\documentclass[aps,pra,twocolumn,showpacs,groupedaddress,amsmath,amssymb]{revtex4-2}
\pdfoutput=1 
\usepackage{amsmath}
\usepackage{amssymb}
\usepackage{graphicx}
\usepackage{bbm}
\usepackage{epsfig}
\usepackage[usenames]{color}
\usepackage{bm}
\usepackage{amsthm}
\usepackage{mathrsfs}
\usepackage{subfigure}
\usepackage[colorlinks, linkcolor=blue,  anchorcolor=blue, urlcolor=blue, citecolor=blue ]{hyperref}

\newcommand{\nn}{\nonumber}

\newcommand{\sgn}{\operatorname{sgn}}
\newcommand{\bk}{\mathbf{k}}

\newcommand{\dg}{\dagger}

\newcommand{\ve} {\varepsilon}

\newcommand{\be}{\begin{eqnarray}}
\newcommand{\ee}{\end{eqnarray}}
\newcommand{\la}{\langle}
\newcommand{\ra}{\rangle}
\newcommand{\rar}{\rightarrow}

\begin{document}
		
		\title{Many-body slow quench dynamics and nonadiabatic characterization of topological phases}
		\author{Rui Wu}
		\altaffiliation[]{These authors contributed equally to this work.}
		\author{Panpan Fang}
		\altaffiliation[]{These authors contributed equally to this work.}
		\author{Chen Sun}
		\author{Fuxiang Li}
		\email[Corresponding author: ]{fuxiangli@hnu.edu.cn}
		\affiliation{School of Physics and Electronics, Hunan University, Changsha 410082, China}
		\date{\today,\now}
	\begin{abstract}

	 Previous studies have shown that the bulk topology of single-particle systems can be captured by the band inversion surface or by the spin inversion surface  emerged on the time-averaged spin polarization. Most of the studies, however, are based on  the single-particle picture even though the systems are fermionic and of multi-bands. Here, we study the many-body quench dynamics of topological systems with all the valence bands fully occupied, and show that the concepts of band inversion surface and spin inversion surface are still valid. More importantly, the many-body quench dynamics is shown to be reduced to a nontrivial three-level Landau-Zener model, which can be solved exactly. Based on the analytical results,  the topological spin texture revealed by the time-averaged spin polarization can be applied to characterize the bulk topology and thus provides a direct comparison for future experiments.
\end{abstract}

\date{\today}
\maketitle 
 
\section{INTRODUCTION}
	
Since the discovery of the integer quantum hall effect in 1980s \cite{1980IQHE}, the concept of topological insulator have ignited extensive interests in condensed matter physics \cite{Laughlin1,Halperin, Haldane}. Although this concept was first proposed in the electronic systems, it has been extended to almost all branches of physics from optics, acoustics, mechanics to magnetics and spintronics \cite{Bernevig,OP1,OP2}. The most exotic property of  topological insulators is that they can support  protected boundary states that are immune to moderate {disorder} or defects \cite{Thouless}. In addition, according to the bulk-boundary correspondence, the number of these boundary states are related to the bulk topological invariant of the topological phases \cite{TI,TIS}. Thus, how to characterize bulk topology of the systems plays an important role in exploring the properties of topological insulator.

 However, in the context of traditional band topology, most of { previous studies} about topological characterization focus on the thermodynamically equilibrium systems, while the topological characterization of far from equilibrium systems is less known in general. In the last two decades, due to the rapid development of technology in quantum manipulation and measurement of ultracold atoms \cite{Ultracoldatoms1,Ultracoldatoms2,Ultracoldatoms3}, the platforms of well-designed optical lattice have emerged in quantum physics, where the systems can undergo a dynamical topological transition by dynamically tuning the Hamiltonian parameters, or coupling the systems to a nonequilibrium bath \cite{Science2021,PRB2019,nphys}. Due to its unique advantages in characterizing the topological properties of systems, there have emerged lots of detection methods of nonequilibrium, such as DQPT (dynamical quantum phase transition) \cite{DQPT1,DQPT2,DQPT3,DQPT4}, ES (entanglement spectrum) \cite{ES1,ES2}, and DWN (dynamical winding number) \cite{DWN}, etc. Notably, Liu and his co-workers proposed a new dynamical characterization theory based on quantum quench dynamics   which has been verified experimentally in Hermitian systems \cite{2018Sci,experiment1, experiment2, experiment3,experiment4,experiment5}. They established a dynamical bulk-surface correspondence, which states a generic $d$D topological phase with ${\mathbb{Z}}$-type invariants can be characterized by the $(d-1)$ D invariants defined on the so-called  band inversion surface (BIS). Based on the higher-order BIS,  they then extend their characterization scheme to other AZ tenfold classes of topological phases including ${\mathbb{Z}_2}$-type invariants. By suddenly quenching the system from a trivial phase to a topological {phase}, the bulk topology of the $d$D equilibrium phases of the post-quench Hamiltonian can be easily determined by the BIS emerged on the time-averaged spin polarization (TASP).  
 
 {Nevertheless}, most of the previous studies are based on  the single-particle picture even though the systems are fermionic and of multi-bands. Nonequilibrium quantum many-body systems  may exhibit novel phenomenons that are completely different from the traditional equilibrium many-body systems \cite{Measurement,Nonequilibrium}. Thus, many fundamental concepts, such as the spontaneous symmetry breaking and the topological order, need to be reexamined in the background of nonequilibrium physics. In the case of dynamical characterization, it would be necessary to investigate the effect of the Pauli exclusion principle when the energy levels are occupied by more than one particles who repel each other.
 	
In this paper, rather than characterizing the single-particle systems, we ask how the dynamical topological characterization should be performed in  many-body fermionic systems by considering {the} Pauli exclusion principle. Specifically, we study the four-band topological systems with two lower bands (valence bands) being fully  occupied.  We show that, in such systems, the dynamics is controlled by an effective $3\times 3$ Hamiltonian. Therefore, when considering the slow quench dynamics starting from the deep trivial phases to the topological phases \cite{slow1,slow2}, one encounters a nontrivial three-state Landau-Zener (LZ) problem.  By investigating the  intrinsic symmetry of scattering matrix and utilizing the integrability conditions \cite{LZ1,LZ2,LZ3,LZ4}, we solve this new LZ model, thus adding a new ingredient into the class of exactly solvable time-dependent models. The time averaged spin polarization can then be obtained analytically as a function of quenching rate. We further obtain the topological spin texture on the position of BIS and spin inversion surface (SIS), and thus the dynamical characterization of bulk topology can be construed. 
	
	The rest of the paper is organized as follows. In {Sec.~\ref{sec2}}, we introduce a $4 \times 4$ Hamiltonian for the single-particle system, and then extend it to a  $6 \times 6$ Hamiltonian in the case with  the valence bands being fully occupied. In Sec.~\ref{sec3}, we consider the sudden quench dynamics and slow nonadiabatic quench dynamics of these systems, respectively. In Sec.~\ref{sec4}, we apply the model Hamiltonian to  the $\mathbb{Z}_2$-type and $\mathbb{Z}$-type topological insulator and present the dynamical characterization under slow nonadiabatic quench dynamics.  Finally, a brief discussion  and {summary} is  represented in Sec.~\ref{sec5}.

	\begin{figure}[t]
		\centering
		\includegraphics[width=3.4 in]{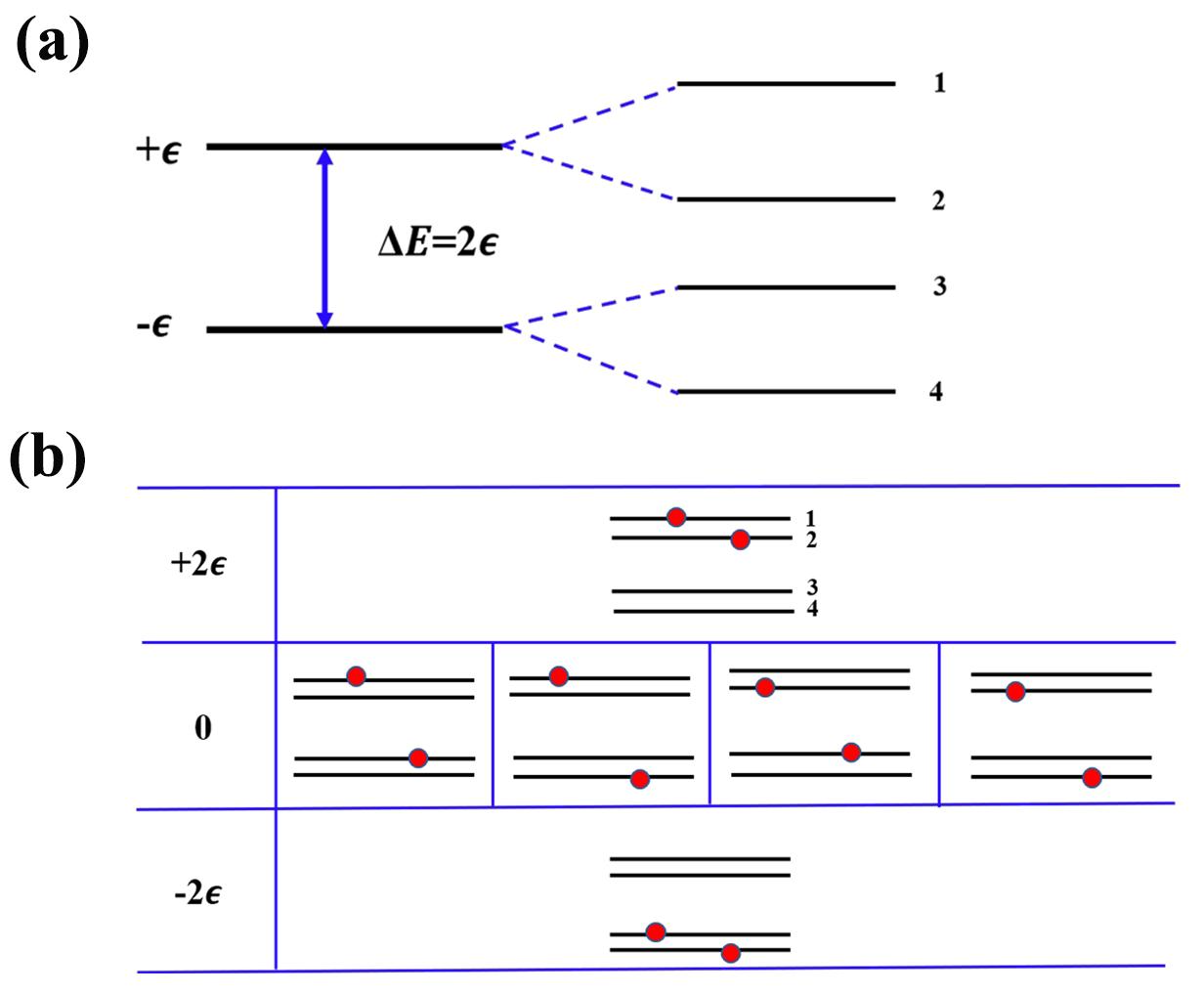}
		\caption{(a) The graphic diagram of the energy levels of the single-particle systems. Here, both the upper and lower bands are doubly degenerate. We denote the sub-band as $ 1,2,3,4$ from top to bottom, respectively. (b) The electron occupation on each energy level in the many-particle system.   The red balls and the black lines represent the electrons and  the energy levels, respectively.}	
		\label{energy_level}	
	\end{figure}

 	 	\section{ Many-body Hamiltonian of the topological system}\label{sec2}
 	We first present the theoretical framework of the many-body version of a four-band topological system  the two valence bands being fully occupied. Specifically, we consider a large class of topological phases whose matrix representations are $4\times4$.
 	The corresponding single-particle Hamiltonian is usually expressed in terms of Gamma matrices:
 	\be
 	H_s(\bk) = \vec{h}(\bk)\cdot \vec{\gamma} = \sum_{j=0}^d h_{j}(\bk) \gamma_{j} \label{Hs}
 \ee
  Here, $\overrightarrow{h}(\bm{k})$ is an effective field which depends on the Bloch momentum $\bm{k}$ in Brillouin zone.  $\gamma_j$ are $4$ by $4$ matrices and obey the Clifford algebra $\{ \gamma_j, \gamma_l\} = 2 \delta_{jl}$ for $j, l = 0, 1, \ldots, d$. {This Hamiltonian can describe the 3D  $\mathbb{Z}$-type insulators and  $\mathbb{Z}_2$ topological insulators, corresponding to $d=3$ and $d=4$, respectively.} Without loss of generality, we choose the following convention for Gamma matrices: $\gamma_0=\sigma_0\otimes \sigma_z$, $\gamma_{1, 2, 3}=\sigma_x\otimes \sigma_{x, y, z}$, $\gamma_4= \sigma_x \otimes \sigma_0$, with $\sigma_{x, y, z}$ the three Pauli matrices and $\sigma_0$ the $2$ by $2$ identity matrix.  In matrix form, the single-particle Hamiltonian $H_s$ can be written explicitly as: 
	\be
 H_s= \left( \begin{array}{cccc}
 		h_0 & 0 & h_- & h_-'  \\
 		0 & h_0 & h_+' & -h_+  \\
 		h_+ & h_-' &-h_0 & 0  \\
 		h_+'  & -h_- & 0 & -h_0
 	\end{array}
 	\right),
 	\label{Hf}
 	\ee
 	  where, for simplicity, we introduce two new parameters: $h_{\pm} = h_1\pm i h_2$, and $h_{\pm}' = h_3\pm i h_4$. 
 	 The eigenvalues of the $H_s$ are $$E_\pm=\pm {\varepsilon}, ~~{\rm  with}~~ {\varepsilon} =\sqrt{\sum_{j=0}^4 h_j^2}.$$ Both of the two eigenvalues are doubly degenerate as shown in Fig.\ref{energy_level} (a).

 	In the nonequilibrium quench dynamics of the topological phases, most of the previous studies only consider  { the single-	particle state \cite{2018Sci,slow1,2021tenfold,PRX2021}}. { However, in the real experiments,  what one should prepare is actually a} many-body state with the two degenerate valence bands fully occupied \cite{experiment1,experiment2,experiment3}. Thus, it is necessary to consider a many-body version of the Hamiltonian with two non-interacting fermions. Note that here we do not specify explicitly the spin degree of freedom, and simply assume that each energy level can host only one fermion due to the Pauli exclusion principle. For this purpose, we will first rewrite the Hamiltonian in terms of creation and annihilation operator for each energy level:  $\hat{H} = \sum_{i, j=1}^4 c^{\dg}_i H_{s, ij} c_j$. For a system with $2$ fermions, the dimension of the Hilbert space would be $6$ with the following basis: 
 	\be
 	|1\ra = c_1^{\dg} c_2^{\dg} |vac\ra,  ~	|2\ra = c_1^{\dg} c_3^{\dg} |vac\ra,  \nn\\
 	|3\ra = c_1^{\dg} c_4^{\dg} |vac\ra,  ~ 	|4\ra = c_2^{\dg} c_3^{\dg} |vac\ra,  \nn \\	
 	|5\ra = c_2^{\dg} c_4^{\dg} |vac\ra, ~  	|6\ra = c_3^{\dg} c_4^{\dg} |vac\ra,   \label{base1}
 	\label{base6} 
 	\ee
 	Here, $c_i$ and $c_i^{\dg}$ denote  the fermionic annihilation and creation operator of band $i$, respectively.  $|vac\ra$ is the vacuum state, which means there are no particles in the system. {As shown in Fig.~\ref{energy_level} (b), the energy levels  of the single-particle and two-particle system are plotted.} In contrast to the four-fold degenerate levels with zero energy, both the levels with $E=+2{\varepsilon}$ and $-2{\varepsilon}$ band are not degenerate. By using the Wick theorem, the elements of Hamiltonian $\hat{H}$ in the many-body basis  $|j\ra$ with $j=1, \ldots, 6$ can be simply obtained, leading to a {$6 \times 6$ }many-body Hamiltonian: 
    		\be
    	H_m=	\left(  \begin{array}{cccccc}
    			
    			2h_0 & h_+'  & -h_+ & -h_- & -h_-' & 0\\
    			h_-' & 0 & 0 & 0 & 0 & -h_-'\\
    			-h_- & 0 & 0 & 0 & 0 & h_-\\
    			-h_+ & 0 & 0 & 0 & 0 & h_+\\
    			-h_+' & 0 & 0 & 0 & 0 & h_+'\\
    			0 & -h_+' & h_+ & h_- & h_-' & -2h_0
    		\end{array}
    		\right). \label{Hm}
    		\ee
\section{Many-body quench dynamics}\label{sec3}
In this section, we will study the many-body quench dynamics of topological systems based on the many-body Hamiltonian $H_m$. We will first study the sudden quench dynamics, in which one parameter of the Hamiltonian is suddenly changed to a new value, and the initial ground state will evolve according to the new Hamiltonian. This protocol has been well studied in the single-particle picture, and has been applied to characterize the topological phases. Then we will study the slow quench dynamics, in which  one parameter is slowly changed to a new value during a finite time interval. This protocol complicates the problem. But it was pointed that slow quench protocol also provides a more efficient method in characterizing the topological phases~\cite{slow1,slow2}. 

\subsection{Sudden quench dynamics}
	For sudden quench, we assume the system stays initially at the ground state and then suddenly change the Hamiltonian. We will choose  $h_0$ as the quench axis. Initially, we assume $h_0$ to be infinitely large compared with other parameters, and then suddenly change $h_0$ to be finite value. Due to the Hamiltonian is time-independent, after the sudden quench, the system will evolve according to the post-quench Hamiltonian. At an arbitrary time $t$, the state evolves into $|\psi(t)\ra=e^{-iH{t }}|\psi(0)\ra=\sum_{j=1}^{6}a_j  e^{-iE_jt}|j\ra$, where $a_j$ is the superposition coefficient of initial wave function projected on to the eigenstates $|j\ra$ of the post-quench Hamiltonian. We will be interested in the experimentally measurable quantity time averaged spin polarization (TASP), which can be calculated directly~\cite{2018Sci}: $$\overline{\la\bm{\gamma}\ra}=\lim_{T\rightarrow\infty}\frac{1}{T}\int_{0}^{T}\la\psi(t)|\bm{\gamma}|\psi(t)\ra dt. $$

	 In the experiment of ultracold atoms, a 2D quantum anomalous Hall  insulator has been realized  \cite{experiment1, experiment3}.By  quenching the Hamiltonian from a trivial to a topological parameter regime, it was observed that a novel ring pattern of the spin dynamics emerges in momentum space during unitary evolution.  The spin oscillation is quantified by  measuring the momentum-dependent spin polarization with the number of atoms in the spin up and down states detected through the spin-resolved time-of-flight imaging.  The resulting dynamical ring pattern can be uniquely identified  to a nontrivial postquench topology. This scheme displays a high precision determination of the full phase diagram for the system's band topology.
	
	We choose the ground state as the initial state $|\psi(0)\ra$, i.e, $|6\ra$. In the meanwhile, the system is quenched from a deep trivial phase to a topological nontrivial phase. Direct calculation gives a simple result of TASP,	
	\be
	\overline{\la\gamma_i\ra}=\frac{2h_i}{{\varepsilon}}(|a_1|^2-|a_6|^2),
	\ee
	where $a_6=\frac{h_0+{\varepsilon}}{2{\varepsilon}}$, $a_1=\frac{h_0-{\varepsilon}}{2{\varepsilon}}$. 	
    Aftering substituting the $a_6$ and $a_1$, we readily  obtain the TASP in sudden quench as follows:
	\be
	\overline{\la\gamma_i\ra}=-\frac{2h_ih_0}{{\varepsilon}^2},	
	\ee 
	The description of the TASP here is similar to Liu's paper, except that the  value is doubled \cite{2018Sci,2021tenfold}. 
	
 	\subsection{Slow quench dynamics} 
 	In slow nonadiadatic quench, the Hamiltonian is time-dependent, i.e, $h_0(t)=h_0+\frac{g}{2t}$ with $g$ determining the quench rate. If $g$ is large, the quench is slow and adiabatic. If $g$ tends to zero, the quench reduces to the case of sudden quench.  When the time $t$ goes from $0^+$ to infinity, the system undergoes a transition from a trivial phase to a topological phase. Thus, to study the slow quench dynamics and calculate the spin polarization, one needs to first solve the corresponding time-dependent Schr\"odinger equation with time-dependent Hamiltonian, i.e., the famous Landau-Zener problem~\cite{slow1,LZ1,LZ2}. Here, we encounter  a six-state Landau-Zener problem.  Exactly solvable multi-state Landau Zener problems must satisfy integrability conditions~\cite{LZ1,LZ2,LZ3,LZ4}. Intriguingly, we find that this $6$-state Landau-Zener problem {can be exactly solvable}, by reducing the $6 \times 6$ Hamiltonian $H_m$ to an equivalent $3 \times 3$ Hamiltonian under an unitary transformation. Specifically, by introducing a unitary matrix  
 \be
 M=\frac{1}{\sqrt{2}h}\left(  \begin{array}{cccccc}
 	\sqrt{2} h  & 0 & 0 & 0 & 0 & 0\\
 	0 & -i h_- & -i h_+ & -i h_+' & ih_-' & 0  \\
 		0 & -h_+' & h_-' & h_- & h_+ & 0\\
 		0 & h_+ & -h_- & h_-' & h_+'  & 0\\
 		0 & -i h_-' & -i h_+' & i h_+& -i h_- & 0\\
 		0 & 0 & 0 & 0 & 0 & \sqrt{2} h
 	\end{array}
 	\right),
 	\nonumber
 	\ee
 with $h=\sqrt{h_1^2 + h_2^2+h_3^2+h_4^2}$, 	one can an obtain a physically equivalent Hamiltonian by an unitary transformation $H_{m}^{(6\times6)}=M H_m M^\dagger$:
 	\be
 	H_{m}^{(6\times6)}=\left(  \begin{array}{cccccc}
 		2h_0 & \sqrt{2}h  & 0 & 0 & 0 & 0\\
 		\sqrt{2}h & 0 & 0 & 0 & 0 & -\sqrt{2}h  \\
 		0 & 0 & 0 & 0 & 0 & 0\\
 		0 & 0 & 0 & 0 & 0 & 0\\
 		0 & 0 & 0 & 0 & 0 & 0\\
 		0 & -\sqrt{2}h & 0 & 0 & 0 & -2h_0	
 	\end{array}
 	\right).
 	\ee	
 	Note that the unitary matrix $M$ is independent of $h_0$. Therefore, this unitary transformation is always valid even for the time-dependent Hamiltonian when $h_0$ is time-independent.
 	
 	One can see that the third, fourth and fifth  levels are fully decoupled from the others, and thus an effectively $3\times 3$ Hamiltonian is obtained. Together with the time dependent part $g/t$ in the $h_0$ terms, we explicitly write down the time-dependent Hamiltonian as: 
  	\be
 	H_m^{(3\times3)}(t)=\left(  \begin{array}{ccc}
 		2h_0 +  g/t & \sqrt{2}h  & 0\\
 		\sqrt{2}h & 0 & -\sqrt{2}h \\
 		0 & -\sqrt{2}h & -2h_0 - g/t\\	
 	\end{array}
 	\right). 
 	\ee
 	One thus needs to solve the $3$-state LZ problem: 
 	\be
 	i\partial_t |\psi(t) \ra = H_m^{(3\times3)}(t) | \psi(t)\ra.  \label{schro}
 	\ee 
  The goal of the multistate Landau–Zener theory is to analytically determine elements of the scattering matrix and the transition probabilities between states of this model after evolution with such a time-dependent Hamiltonian from initial time to positive infinite time.  To analytically solve such a time-dependent model is a nontrivial task. Nevertheless, the results would serve as very useful tools in studying the quantum dynamics of driven systems, and provide insights in dynamical phenomena such as Kibble-Zurek mechanism \cite{KZ}, quantum annealing \cite{annealing} and so on. However, the exactly solvable models are rare and have been found to exist only in some special forms such as the bow-tie model \cite{bow-tie} and time-dependent Tavis-Cumming model \cite{QED}. We will solve the three-state LZ problem (\ref{schro})  by investigating the intrinsic symmetry and applying the integrability constraints on the scattering amplitudes. We present the results here and leave the details in the Appendix.

 	We note that, at initial time $t=0^+$, the time-dependent term is infinitely large and thus the ground state eigenvector is $|\psi_i\ra = (0, 0, 1)^T$. At final time $t\rar \infty$, the eigenvalues are 
 	\be
 	E=\pm2{\varepsilon}, 0,
 	\ee
 	with $\varepsilon = \sqrt{h_0^2 + h^2} = \sqrt{\sum_{j=0}^4 h_j^2}$,
 	and the corresponding instantaneous eigenvectors are:
 	\be
 	|\psi_+\ra=\frac{1}{2{\varepsilon}}\left({\varepsilon}+h_0, \sqrt{2}h, h_0-{\varepsilon} \right)^T,\label{eq:psi+}
 	\ee
 	\be
 	|\psi_0\ra=\frac{1}{\sqrt{2}{\varepsilon}}\left(h, -\sqrt{2}h_0, h \right)^T,
 	\ee	
 	\be
 	|\psi_-\ra=\frac{1}{2{\varepsilon}}\left(h_0-{\varepsilon}, \sqrt{2}h, h_0+{\varepsilon} \right)^T.\label{eq:psi-}
 	\ee	
 	\begin{figure}[t]
 		\centering
 		\includegraphics[width=3.4 in]{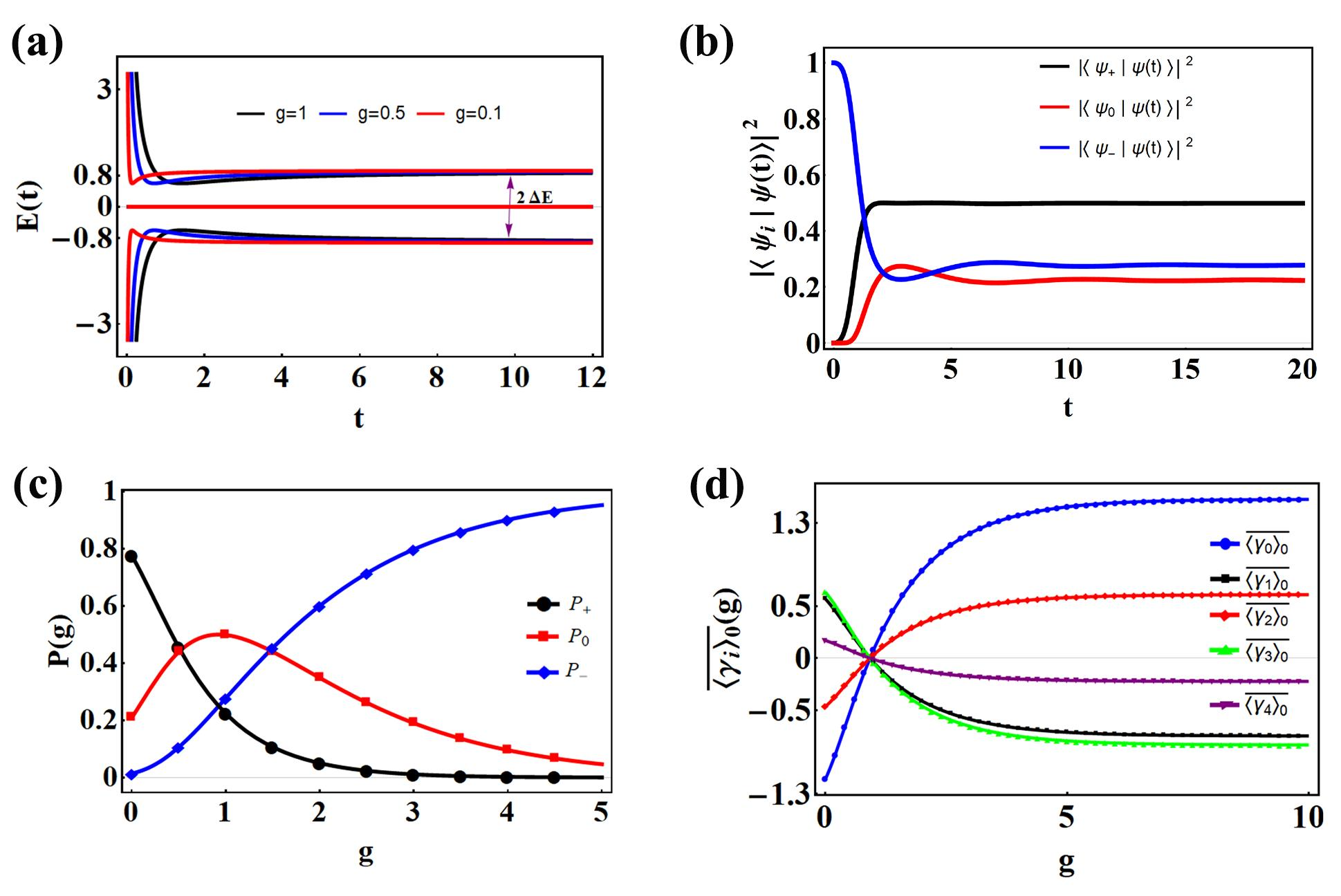}
 		\caption{ (a) The instantaneous eigenenegy as a function of time $t$ for different values of $g$.Here, we choose the point $(k_x, k_y)= (\frac{2\pi}{3}, -\frac{\pi}{4}) $ in the model (\ref{eqZ2}) with parameters $ m=-0.5t_0$, $t_0'=0.5t_0$, $t_{so}=0.2t_0$. (b) Occupation probability of the state vector $|\psi(t)\ra$ on three instantaneous eigenstates with $g=1$. (c) Transition probability $P$ from initial ground state to final state as a function of $g$. These colored points and solid lines are numerical results and analytic results, respectively. The analytic results are given by Eqs.(\ref{psi-} - \ref{psi-}). In the numerics, time $t$ is taken from 0.0001 to 20000. (d) TASP as functions of varying $g$. }	
 		\label{Et}	
 	\end{figure}
 	 	{Taking the $\mathbb{Z}_2$-type topological insulator (discussed below) as an example, we plot the instantaneous energy spectrum of $H_m^{(3\times3)}$ in Fig.~\ref{Et}(a)}. If the initial state is prepared in its ground state $|\psi_i\ra = (0,0,1)^{T}$ at $t\to 0^+$, the system will undergo a nonadiabatic transition during time evolution governed by the time-dependent Schr\"odinger-equation (\ref{schro}). The process of this nonadiabatic transition is illustrated in Fig.~\ref{Et}(b), showing the occupation probability of the $|\psi(t)\ra$ on three instantaneous eigenvectors as a function of time $t$. Starting from the initial state  $|\psi_i\ra$ with unit probability,   the energy gap of the system gradually becomes smaller, and in the meanwhile, the 
 	 occupation probability of $|\psi(t)\ra$ on instantaneous ground eigenvectors decreases, while the probability on the other two instantaneous  eigenvectors increases. At the long time limit, the energy gap remains as a constant and  corresponding  occupation probability on each energy level arrives its saturation value.	 

 	The transition probability of the initial ground state $|\psi_i\ra=(0, 0, 1)^T$  to the final three eigenvectors $ |\psi_-\ra$, $ |\psi_0\ra$ and $ |\psi_+\ra$ at $t\to \infty$ is given as, respectively: 	
 \be
 P_{-}&=&\Big(\frac{e^{\pi g}-e^{-\pi g_c}}{e^{\pi g}-e^{-\pi g}}\Big)^2
 \label{psi-}, \\
 P_{0}&=&2\frac{(e^{\pi (g-g_c)}-1)(1-e^{-\pi (g+g_c)})}{(e^{\pi g}-e^{-\pi g})^2},
 \label{psi0} \\
  P_{+}&=&\Big(\frac{e^{-\pi g_c}-e^{-\pi g}}{e^{\pi g}-e^{-\pi g}}\Big)^2
 \label{psi+}, 
 \ee
 with
 \be
 g_c=g\frac{h_0}{\varepsilon},
 \ee 
Note that the unitary condition is preserved $P_{-}+P_{+}+P_{0}=1$. 
 	
 	 One can see the system experiences a transition from nonadiabatic evolution to adiabatic evolution in Fig.~\ref{Et}(c) while varying $g$ from 0 to $\infty$. 
 Based on the above  Landau-Zener probability, one can write down the final state vector at long-time limit,
 		\be	
 		|\psi(t)\ra&=&\sqrt{P_{+}}e^{-i2{\varepsilon} t+\delta_1}|\psi_+\ra+\sqrt{P_{-}}e^{i2{\varepsilon} t+\delta_2}|\psi_-\ra  \nn \\
 		&&+\sqrt{P_{0}}|\psi_0\ra ,	
 		\label{psit}	
 		\ee
 where  $\delta_1$ and $\delta_2$ are two undetermined relative phase factors.
 
  Furthermore, one can study the dynamics of spin polarization $\langle \bm \gamma({t})\rangle=\langle\psi(t)|\bm \gamma |\psi(t)\rangle$ and its average (TASP) after a long time period of evolution:
 	\be
 	\overline{\la\bm{\gamma}_i\ra}
 	=\frac{2h_i}{{\varepsilon}}(P_{+}-P_{-}).
 	\label{eq:LZ}
 	\ee 
 	This result agrees well with the results of numerical simulation as shown in Fig.~\ref{Et}(d). When $g \to$ 0, this solution is also the TASP of sudden quench. Here, we define the position $(P_{+}-P_{-})=0$ as SIS \cite{slow1,slow2}.
 
 	Before discussing the specific topological models, it should be mentioned that the nonadiabatic protocol is readily accessible in ultracold atomic experiments by using a bias magnetic field to simulate the time-changing part of the quenching axis \cite{experiment1, experiment3}. Furthermore, the spin polarization can be measured by using the spin-resolved time-of-flight imaging.  Moreover, the nonadiabatic transition time $t_c$ can be estimated as $t_c\sim \hbar/\Delta$, which is of the order of $10~ \mu$s. Here,  $\Delta$ is the the  characteristic energy scale.   In order to observe the effect of the nonadiabatic transition in experiment, the nonadiabatic transition time $t_c$ must be much smaller than the decoherence time $\tau$ of the ultracold atomic system, which is longer than $1$ ms. Therefore, this condition is satisfied and the nonadiabatic quench protocol is quite promising in detecting the topological phases in ultracold atoms.

 	 \begin{figure}[t]
 	 	\includegraphics[width=1\linewidth]{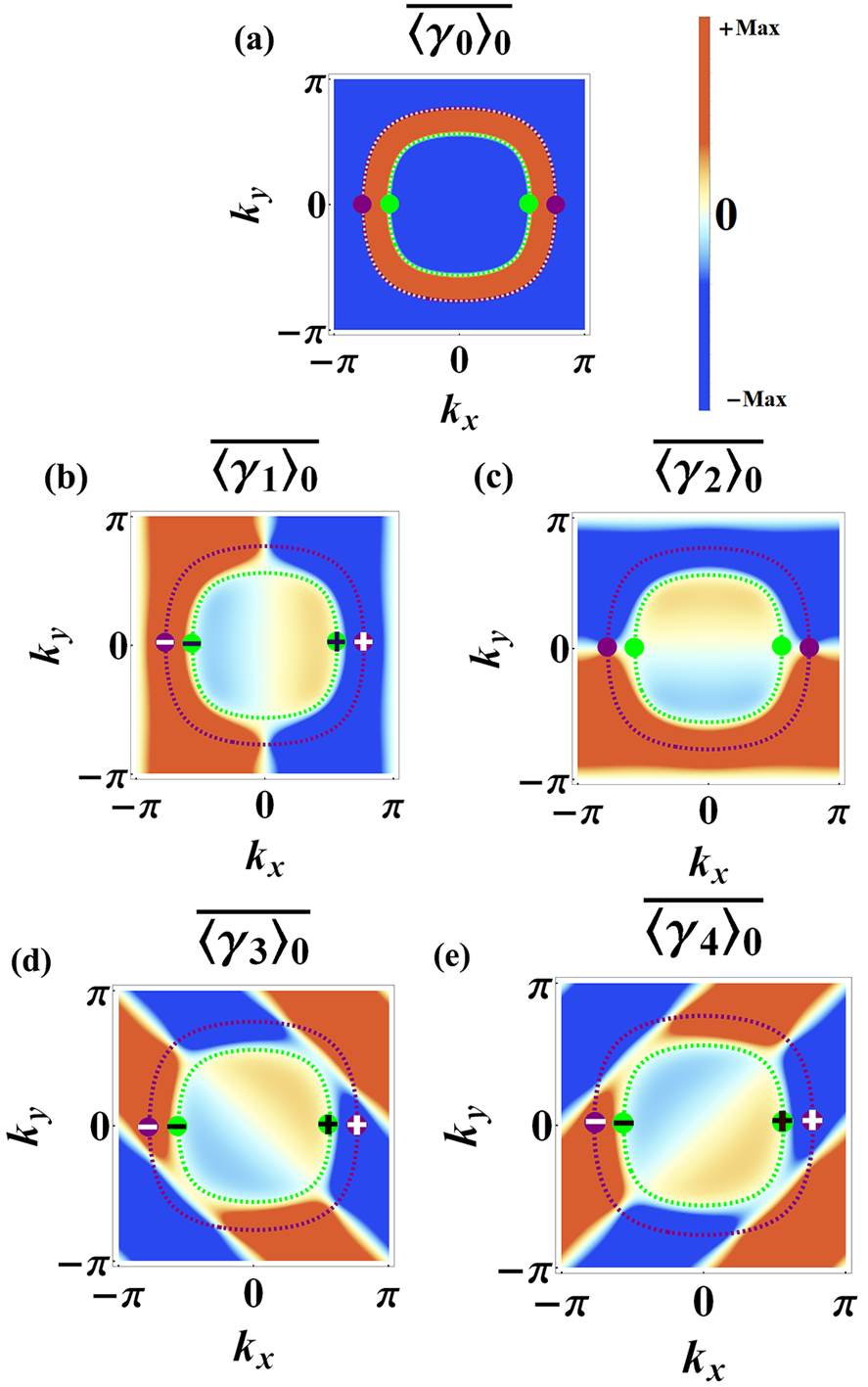}
 	 	\caption{Analytical results of TASP {under slow quench dynamics}. The system is quenched from $t=0$ to $t=\infty$ with $5t_{so}=2t_0'=-2m=t_{0}=1,g=5$. After quenching $h_0$, five components of TASP are shown. The 3-SIS and  3-BIS are denoted by purple dashed lines and  green dashed lines, respectively. The opposite signs on 4-BIS (4-SIS) constructed by $\overline{\la { \gamma}_{0,2}\ra}$ in $\overline{\la { \gamma}_{1,3,4}\ra}$  characterizes the nontrivial topology, and implies the $\mathbb{Z}_2$ invariant $\nu^{(2)}$=$-1$.}	
 	 	\label{slow}			
 	 \end{figure}
	   
 \section{ The nonadiabatic characterization of topological phases}\label{sec4}
 
 Now, we can apply the obtained TASP to the dynamical topological characterization of the real models. The Hamiltonian $H_s$  which we considered in this paper can describe at least two different topological phases. When $h_4 \neq 0$, $H_s$ corresponds to the $\mathbb{Z}_2$ topological insulators, including the 3D first descendant and 2D second descendant insulators. When $h_4 = 0$, $H_f$  corresponds to the 3D chiral topological insulators. {The dynamical characterization of the 3D chiral topological insulators only needs the introduction of 1-BIS and 1-SIS (low-order), while the characterization of the $\mathbb{Z}_2$ topological insulators need the introduction of high-order BIS and high-order SIS \cite{slow2}}. 
 
 \subsection{$\mathbb{Z}_2$-type topological insulator}

 As an example of characterizing $\mathbb{Z}_2$ topological insulator, a 2D  $\mathbb{Z}_2$ topological phase \cite{2021tenfold}  is considered with Hamiltonian: 
 \be
&& h_0=\frac{g}{2t}+m-t_0 \sum_{i=x,y} \cos k_i  \nn \\
 &&~~~~~~~  - t_0' \sum_{i=1,2} \cos [k_x-(-1)^i k_y],\nonumber\\ 
 &&h_{1,2}=t_{so}\sin k_{x,y}, \label{eqZ2}\\
 &&h_{3,4}=t_{so}\sin [k_x\pm k_y],\nonumber
 \ee
 where  $t_0,t_0'$ represent the spin-conserved hopping coefficients while $t_{so}$ denote the spin-flipped hopping coefficients.  The equilibrium topology of this model is characterized by the value of $m$ with the Fu-Kane invariant FK= $\sgn\{(m+2t_0')^2[(m-2t_0')^2-4t_0^2]\}$ \cite{FK1,FK2}.

 \begin{figure}[h]
	\includegraphics[width=200pt]{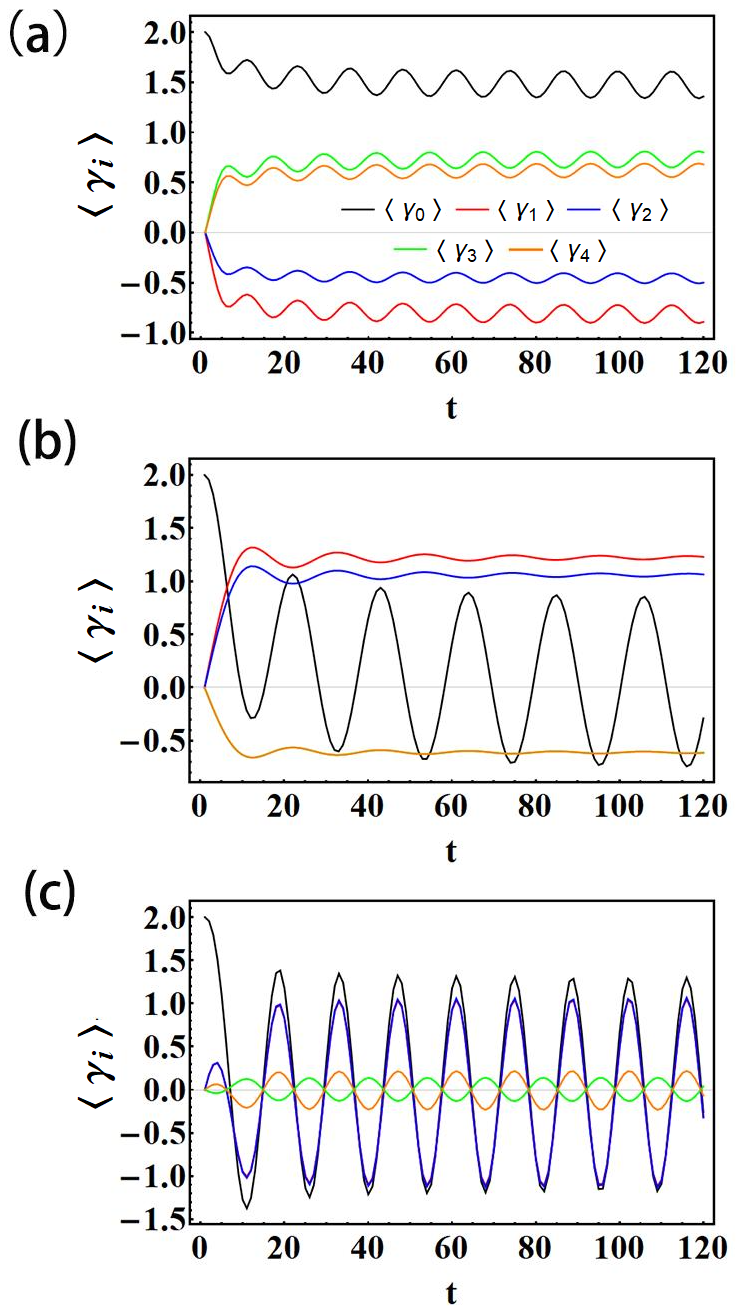}
	\caption{Analytical results of time evolution of spin polarization for 2D $\mathbb{Z}_2$ topological model after slow quenching $h_0$ axis with $t_{so} = 0.2$, $m = -0.5$, $t_0'=0.5$, $t_0$ = $g$ = 1. (a)(b)(c) the corresponding time evolution of spin polarization on a general point (a), BIS (b) and SIS (c), respectively. In (b), ${\la { \gamma}_{3}\ra}$ overlaps with ${\la { \gamma}_{4}\ra}$. In (c), ${\la { \gamma}_{1}\ra}$ overlaps with ${\la { \gamma}_{2}\ra}$. }	
	\label{Gammat}			
\end{figure}

 By slowly quenching $h_0$ from $t=0$ to $t=\infty$ with $m=-0.5$, five components of TASP are observed as shown in Fig.~\ref{slow}. As one can see, 1D 3-BIS and 1D 3-SIS with vanished spin polarization are observed in $\overline{\la { \gamma}_{0}\ra}$, and their corresponding position in other components of TASP are  denoted by purple dashed ring and green dashed ring, respectively. To give an intuitive comprehension of the behaviour of BIS and SIS, we also plot the time evolution of five components ${\la { \gamma}_{0,1,2,3,4}\ra}$ on BIS, SIS and other point in Fig. \ref{Gammat}. One can observe an approximately stable dynamic precession behavior of spin polarization on all three points after a short period of evolution. However, for BIS, there is only  ${\la { \gamma}_{0}\ra}$ oscillating between positive and negative value. Thus, the long-time averaged value of ${\la { \gamma}_{0}\ra}$ will vanish, while other components of TASP {are} still nonzero. Compared with BIS, ${\la { \gamma}_{0,1,2,3,4}\ra}$ all oscillates {on SIS} between positive and negative value, corresponding to a vanishing spin polarization in all components of TASP.  {These features} can be convenient to identify the position of SIS and BIS based on TASP in experiments. In addition, all the components of spin polarization of a general point do not oscillate between positive and negative value, and thus have no distinguished {features} in TASP.
 
 Moreover, the 0D 4-BIS and 0D 4-SIS can be easily constructed on 3-SIS and 3-BIS by the intersection of vanishing polarization between two components of TASP. Here we choose  $\overline{\la { \gamma}_{0}\ra}=0\cap\overline{\la { \gamma}_{2} \ra}=0$ as an example. As shown in Fig.~\ref{slow}(a) and (c),  the 0D 4-BIS and 0D 4-SIS {are} indicated by the purple points and green points, respectively. The  values of 0D BIS and gradients of 0D SIS are indicated by the opposite signs as shown in Fig.~\ref{slow} (b), (d) and (e), implying the nontrivial  $\mathbb{Z}_2$ index $\nu^{(2)}$=$-1$.

 	\subsection{$\mathbb{Z}$-type topological insulator } 
 	We proceed to give the characterization scheme of the high-dimensional  $\mathbb{Z}$-type topological phases under slow  nonadiabatic quench dynamics. The Hamiltonian of 3D topological phase can be written as $H(k)=\sum_{i=0}^{4}h_i\gamma_i$, with
 	\be
 	&& h_0=\frac{g}{2t}+m-t_0\sum_{j=x,y,z}\cos k_i,\nonumber\\
 	&& h_{1,2,3}=t_{so}\sin k_{x,y,z},\nonumber\\
 	&& h_4=0.
 	\ee
 	We quench this system from $t=0^+$ to $t\rightarrow\infty$ with large $m$ to a finite value, and thus the system finally lies in different phases depending on the value of $m$: for $|m|>3t_0$, the trivial phase; for $t_0<m<3t_0$, the topological phase with winding number $\nu_3=-1$; for $-t_0<m<t_0$ with $\nu_3=2$, and for $-3t_0<m<-t_0$, with $\nu_3=-1$.

\begin{figure}[t]
 		\includegraphics[width=3.4in]{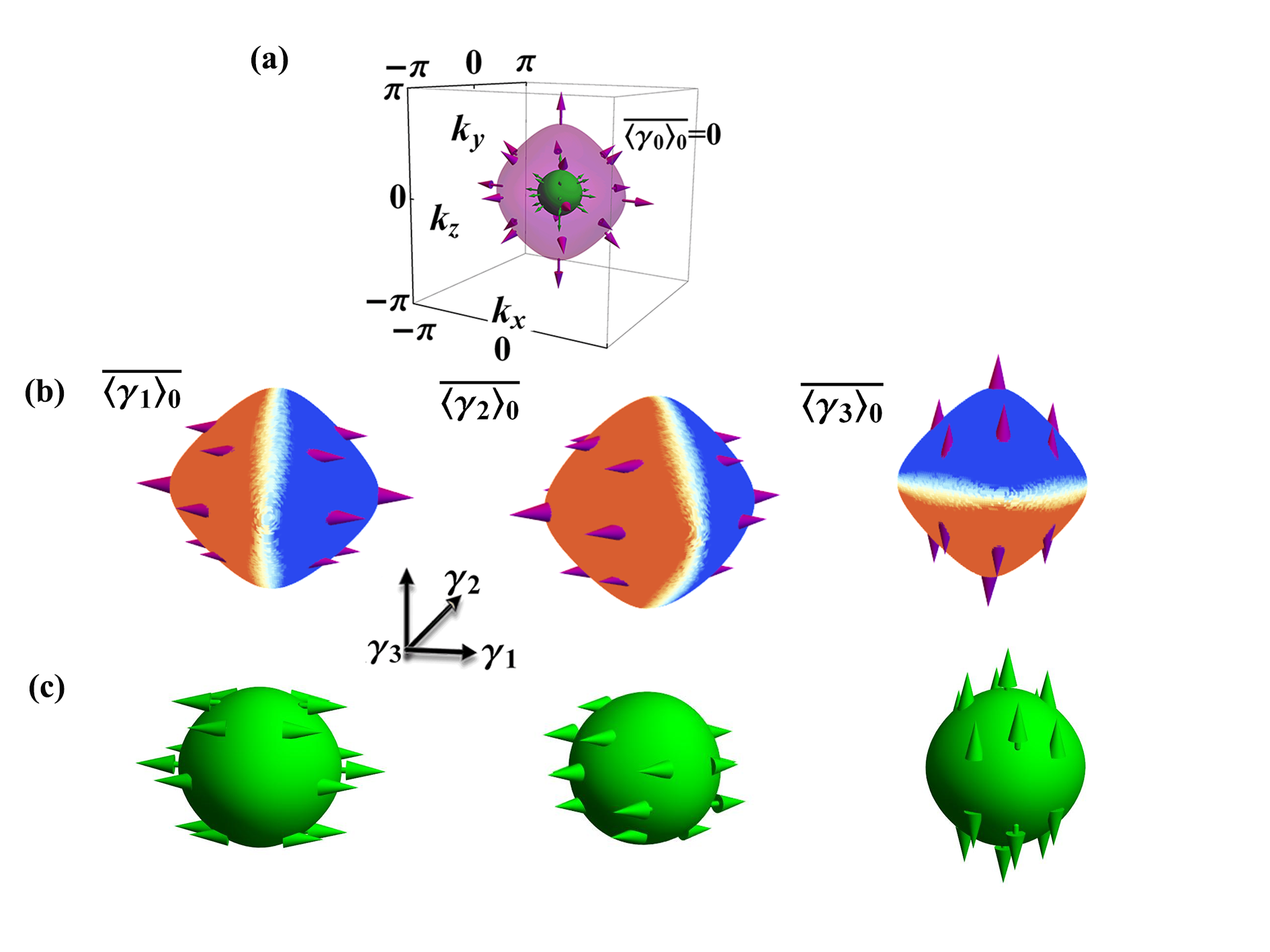}
 		\caption{Analytical results of TASP for the 3D chiral topological insulator under slow quench dynamics. Here we set $t_{so}=t_0=g=1$, $m=1.5$.  In (a), the purple surface is 1-BIS which is defined by $\overline{\la\gamma_{0}\ra}=0$, and the purple arrows are the dynamical field shown in (b). The green one is SIS, and the green arrows are the dynamical field  shown in (c). In (b) and (c), the three components of TASP $\overline{\la\gamma_{1,2,3}\ra}$ on 1-BIS and 1-SIS are shown. The dynamical field $-\overline{\la\gamma_{1,2,3}\ra}$ and $\widetilde{g_{1,2,3}}$ are illustrated by purple and green arrows, respectively. }	
 		\label{3Dchiral}	
 	\end{figure}

 	As shown in Fig.\ref{3Dchiral} (a), one can see both BIS and SIS (purple and green surfaces) become a closed surfaces where spin polarization vanished in $\overline{\la\gamma_{0}\ra}=0$. In Fig.\ref{3Dchiral}(b) and (c), the non-zero values of spin polarization $-\overline{\la\gamma_{1}\ra}$, $-\overline{\la\gamma_{2}\ra}$, $-\overline{\la\gamma_{3}\ra}$ are plotted by the purple and green arrows and combined on the BIS in $\overline{\la\gamma_{0}\ra}$. The above combined purple arrows display a three-dimensional topological pattern on the BIS surface, and thus give a nontrivial winding number $\nu_3=1$. In addition, the topological characterization of SIS is the same as that of BIS, except that the dynamical field $-\overline{\la\gamma_{1,2,3}\ra}$  is replaced by gradient  $\widetilde{g_{1,2,3}}$.

 	\section{DISCUSSION and CONCLUSION}\label{sec5}

 	 In summary, we have studied the many-body dynamical characterization of topological phases with a focus on whether the Pauli exclusion principle would affect the transition probability after slow quench dynamics. By investigating the TASP of many-body  fermionic systems under nonadiabatic quench dynamics, we find that the TASP of two-particle is the algebraic sum of that of the single-particle system. Albeit simple, this result is quite nontrivial  as considering the Pauli exclusion principle so that, in the time-evolution, different particles would effectively repel each other. Moreover, to achieve the results, we have obtained a new three-level Landau-Zener model which adds to the class of exactly solvable models. This nontrivial model is solved by investigating the intrinsic symmetry of the time-dependent Hamiltonian and by applying the integrability conditions. The obtained TASP (also
 	confirmed numerically) can be applied to characterize the bulk topology of system in nonequilibrium state. Furthermore, our results suggest a general method for exploring the band topology of noninteracting many-body fermionic systems. Particularly, it is worthwhile to notice that the nonadiabatic quench can be realized in ultracold atomic experiments because the nonadiabatic transition time $t_c$ much smaller than the decoherence time of the ultracold atomic system $\tau$. Thus, one can expect our results may provide reference in future experiment of (super)weak interaction.
\section*{Acknowledgements}
This work was supported by the National Key Research and Development Program of Ministry of Science and Technology (No. 2021YFA1200700), National Natural Science Foundation of China (No. 11905054, No. 12275075 and No.12105094) and  the Fundamental Research Funds for the Central Universities from China.

\section*{Appendix}
In the appendix, we provide the details of solving the new three-state Landau-Zener model. Explicitly, the time-dependent Schr\"odinger equation is: 
	\be
 	i\frac{d}{dt} |\psi(t) \ra = H(t) | \psi(t)\ra,  \label{schro}
 	\ee 
 	with
	\be
 	H(t)=\left(  \begin{array}{ccc}
 		2h_0 +  g/t & \sqrt{2}h  & 0\\
 		\sqrt{2}h & 0 & -\sqrt{2}h \\
 		0 & -\sqrt{2}h & -2h_0 - g/t\\	
 	\end{array}
 	\right). 
 	\ee
The time $t$ is from $0^+$ to $\infty$. At initial time, the instantaneous eigenenergies are $\infty$, $0$, and $-\infty$, which we denote as $1$, $2$ and $3$ levels. At final time $t= \infty$, the instantaneous eigenenergies are  $\pm 2\ve$ and $0$ with eigenvectors given in Eqs.~(\ref{eq:psi+})-(\ref{eq:psi-}),  which we denote as $\pm$, and $0$ levels, respectively. We define transition probability $P_{\mu j}$ and the corresponding scattering matrix $\hat{S}_+$ with elements $\hat{S}_{\mu j}$ from initial $j$-th level to final $\mu$-th level, with $j=1, 2, 3$, and $\mu=+, 0, -$. Clearly, one has $P_{\mu j} = | \hat{S}_{\mu j}|^2$.  
We explicitly write down the scattering matrix \cite{Sinitsyn2014, Sinitsyn2017}:
\be
\hat{S}_+ = \left(  \begin{array}{ccc}
 		s_{+1} & s_{+2}  & s_{+3}\\
 		s_{01} & s_{02}  & s_{03} \\
 		s_{-1} & s_{-2}  & s_{-3}\\	
 	\end{array}
 	\right). \label{matrixS1}
\ee
Note that the Schr\"odinger equation (\ref{schro}) possess a chiral symmetry. If we write $\psi(t) = (a_1, a_2, a_3)^T$,  Eq.~(\ref{schro}) is invariant when  exchanging $a_1$ and $a_3$ together with taking the conjugate so that $i \rar -i$, or $\hat{H} \rar -\hat{H}$. This joint operation produces a new scattering matrix $\hat{S}'_+$. On one hand, the elements of $\hat{S}'_+$ can be obtained from $\hat{S}_+$ by exchanging the index $1\rar 3$ and $+\rar -$, i,e,:
\be
\hat{S}'_+ = \left(  \begin{array}{ccc}
 		s_{-3} & s_{-2}  & s_{-1}\\
 		s_{03} & s_{02}  & s_{01} \\
 		s_{+3} & s_{+2}  & s_{+1}\\	
 	\end{array}
 	\right).
\ee
On the other hand, the two scattering matrices are related with each other through a conjugate operation: $\hat{S}'_+= \hat{S}_+$. Therefore we have several identities: $s_{-3}=s_{+1}^*$, $s_{-2}=s_{+2}^*$, $s_{+3} = s_{-1}^*$, $s_{03}= s_{01}^*$ and $s_{02}=s_{02}^*$. Explicitly, these identities reduce Eq.~(\ref{matrixS1}) to be:
\be
\hat{S}_+ = \left(  \begin{array}{ccc}
 		s_{+1} & s_{+2}  & s_{-1}^*\\
 		s_{01} & s_{02}  & s_{01}^* \\
 		s_{-1} & s_{+2}^*  & s_{+1}^*\\	
 	\end{array}
 	\right). \label{matrixS3}
\ee
Also, $s_{02}$ is real. 

We also have the unitary condition for $\hat{S}_+$: $\hat{S}_+^{\dg} \hat{S}_+ =\hat{S}_+^{\dg} \hat{S}_+ = \hat{1} $, which gives rise to more conditions: 
\be
&&|s_{+1}|^2 + |s_{+2}|^2 + |s_{-1}|^2 =1,  \label{s1}\\
&& 2|s_{01}|^2 + |s_{02}|^2=1, \label{s2} \\\
&& 2|s_{12}|^2 + |s_{02}|^2=1, \label{s3}\\
&&2 s_{+1} s_{-1}^* +s_{+2}^2 =0.  \label{s4}
\ee
From (\ref{s2}) and (\ref{s3}), one obtains $|s_{01}| = |s_{12}|$. From (\ref{s4}) one obtains $|s_{+2}|^2 = 2 |s_{+1}| |s_{-1}|$.  Substituting this relation into (\ref{s1}), one can get $|s_{+1}| + |s_{-1}|=1$. Returning to the transition probability:
\be
\hat{P} = \left(  \begin{array}{ccc}
 		P_{+1} & P_{01}  & P_{-1}\\
 		P_{01} & P_{02}  & P_{01} \\
 		P_{-1} & P_{01}  & P_{+1}\\	
 	\end{array}
 	 	\right),  \label{eqPsym}
\ee
 with relations: 
\be
&&P_{+1} + P_{01} + P_{-1} =1, \\
&& 2P _{01} + P_{02} = 1, \\
&&\sqrt{P_{+1}} + \sqrt{P_{-1}} =1.
\ee
Now we have $4$ parameters to be determined, but only $3$ conditions. One more condition is yet to be established.

We resort to  the connection formula \cite{Sinitsyn2017}. First we make an unitary transformation to the Hamiltonian such that, in the new basis, the Hamiltonian has a form of $\hat{A} + \hat{C}/t$ with the constant matrix $\hat{A}$ being  diagonal. The unitary matrix is 
\be
\hat{V} =  \frac{1}{2 \ve} \left(  \begin{array}{ccc}
 		h_0+ \ve & \sqrt{2}h  & h_0- \ve\\
 		\sqrt{2} h & -2 h_0  & \sqrt{2} h \\
 		h_0- \ve & \sqrt{2}h  & h_0+ \ve\\	
 	\end{array}
 	 	\right). 
\ee
The new Hamiltonian $H'= \hat{V} H(t) \hat{V}^{\dg}$ is 
\be
H' =   \left(  \begin{array}{ccc}
 		2 \ve + g_0 /t & g_h/t  & 0\\
 		g_h/t  & 0  & - g_h/t \\
 		0 & -g_h/t  & -2 \ve\\	
 	\end{array}
 	 	\right),
\ee
with $g_0 = g h_0/\ve$, $g_h = g h /\sqrt{2} \ve$. It has been pointed out that, for a Schr\"odinger equation like the form: 
\be
i \partial_t \psi = (\hat{A} + \hat{C} /t) \psi
\ee
it is equivalent to a Schr\"odinger equation like: 
\be
i\partial_{\tau} \psi = (\hat{A} \tau + 2\hat{C}/\tau) \psi \label{btH}
\ee
which can be obtained by a change of variable: $t= \tau^2/2$.   For Eq.~({\ref{btH}}), the equation is symmetry under time reversal: $\tau\rar -\tau$.  Denoting the scattering matrix of Eq.~(\ref{btH}) from $\tau=0^+$ to $\infty$ as $\hat{S}_+$, and the scattering matrix from $\tau=-\infty$ to $0^-$ as $\hat{S}_-$, the time reversal symmetric guarantees that $\hat{S}_- = \hat{S}_+^{\dg}$. In the complex plane of time, one can draw a semicircle ${\cal C}_+$ with infinite radius in the upper plane connecting the time $\tau=-\infty$ to $\infty$.  Along this contour ${\cal C}_+$, the element $S_{11}^{\rm up}$ in the scattering matrix   is given as $S_{11}^{\rm up}   = e^{-\pi g_0}$. This scattering element should be equal to the one along the  real axis $t\in (-\infty, 0^-)$ and $t\in (0^+, \infty)$ connected by a small semicircle ${\cal C}_0$ around the original point, i.e., $[\hat{S}_+ \hat{S}_0^{\rm up} \hat{S_-}]_{11} = S_{11}^{\rm up}$. Here, $ \hat{S}_0^{\rm up}$ is the scattering matrix of the small contour ${\cal C}_0$, and `up' denotes that  contour ${\cal C}_0$ is in the upper plane. This connection formula gives the following identity\cite{Sinitsyn2017}: 
\be
e^{-2\pi g} P_{+1} + P_{01} + e^{2\pi g} P_{-1} =e^{-2\pi g h_0 /\ve}.
\ee

Now we have four parameters to be determined in (\ref{eqPsym}) and also four identities. Therefore the transition probability can be determined unambiguously. Explicitly: 
 \be
 P_{+1}&=&\Big(\frac{e^{\pi g}-e^{-\pi g_c}}{e^{\pi g}-e^{-\pi g}}\Big)^2
 \label{psi-2}, 
 \ee
 \be
 P_{01}&=&2\frac{(e^{\pi (g-g_c)}-1)(1-e^{-\pi (g+g_c)})}{(e^{\pi g}-e^{-\pi g})^2},
 \label{psi02} \ee
 \be
  P_{-1}&=&\Big(\frac{e^{-\pi g_c}-e^{-\pi g}}{e^{\pi g}-e^{-\pi g}}\Big)^2
 \label{psi+2},   \ee
 \be
 P_{02}&=&\Big(\frac{e^{\pi g}+e^{-\pi g}-2e^{-\pi g_c}}{e^{\pi g}-e^{-\pi g}}\Big)^2,
 \ee
 with
 \be
 g_c=g\frac{h_0}{\varepsilon}.
 \ee

 \end{document}